\documentclass[12pt]{article}

\catcode`\@=11

\global\arraycolsep=2pt
\usepackage{amsbsy,amssymb,latexsym,amsfonts,amsmath}
\usepackage{graphicx,color}

\allowdisplaybreaks

\begin{document}

\begin{flushright}
YITP-25-23
\parbox{4.2cm}

\end{flushright}

\vspace*{0.7cm}

\begin{center}
{ \Large Is chiral supersymmetry emanant or emergent?}
\vspace*{1.5cm}\\
{Yu Nakayama}
\end{center}
\vspace*{1.0cm}
\begin{center}

Yukawa Institute for Theoretical Physics,
Kyoto University, Kitashirakawa Oiwakecho, Sakyo-ku, Kyoto 606-8502, Japan

\vspace{3.8cm}
\end{center}

\begin{abstract}
The extended Majorana Nicolai model is one of the simplest models of supersymmetry realized on a fermionic chain in $1+1$ dimensions. Within a certain parameter region, the extended theory breaks the supersymmetry spontaneously, but it has a distinguished feature that the general counting rule of the Nambu-Goldstone mode does not apply: we observe twice as many low-energy degrees of freedom than the broken lattice symmetry. We argue that the extra degrees of freedom originate from the spontaneous breaking of the emergent chiral supersymmetry. This chiral supersymmetry becomes an emanant symmetry in the non-interacting limit. 
\end{abstract}

\thispagestyle{empty} 

\setcounter{page}{0}

\newpage


\section{Introduction}

The realization of a fermion with a chiral symmetry, in particular with an 't Hooft anomaly, on a lattice is a challenging but important problem. This is because the notion of chirality is tied up with a continuum of space-time, and on-site symmetry cannot realize the chiral symmetry in a simple manner. Sometimes it is just emergent by fine-tuning microscopic parameters in the lattice Hamiltonian while taking the continuum limit. Sometimes, it is emanant \cite{Seiberg:2023cdc}\cite{Seiberg:2024gek}\cite{Chatterjee:2024gje}, meaning that there is a (not necessarily on-site) lattice symmetry that gives rise to the chiral symmetry in the continuum limit.

In this paper, we focus on the (spontaneously broken) chiral supersymmetry in $1+1$ dimensions. There exists a class of theories where supersymmetry is realized only by fermionic degrees of freedom on a lattice. Among them, we can engineer the supercharge so that the ground state breaks the supersymmetry spontaneously. The models studied in \cite{Sannomiya:2016wlz}\cite{Sannomiya:2017foz} have a linear dispersion relation for the Nambu-Godstone fermion (sometimes called Goldstino) and the low-energy degrees of freedom are given by relativistic massless Majorana fermion or Dirac fermion for $\mathcal{N}=1$ or $\mathcal{N}=2$ supersymmetry respectively (see also \cite{Moriya:2018fgr}\cite{Rahmani:2018uue}\cite{Katsura:2022xkg}\cite{Miura:2023hms}\cite{Miura:2023qph}\cite{Miura:2024hnh} for related studies).\footnote{Spontaneously broken supersymmetry in condensed matter physics is discussed in various contexts: see e.g. \cite{Yu:2007xb}\cite{Grover:2013rc}\cite{Blaizot:2015wba}\cite{Blaizot:2017bhs}\cite{Marra:2021xqk}\cite{Ma:2021dua}\cite{Hirokawa:2022ked}\cite{Marra:2023fgh}\cite{Zhang:2024pmt} for recent studies.}

This fact, however, leaves a puzzle: the low-energy degrees of freedom given there are those corresponding to the spontaneous breaking of $\mathcal{N}=(1,1)$ or $\mathcal{N}=(2,2)$ supersymmetry. What is the origin of the doubled degrees of freedom? What was wrong with the general counting rule of the Nambu-Goldstone mode?\footnote{We further note that there is an extra subtly in assigning the dispersion relation to the Nambu-Goldstone mode. The fermionic Nambu-Goldstone mode violates the Chadha-Nielsen rule \cite{Nielsen:1975hm} (see also  \cite{Watanabe:2012hr}\cite{Hidaka:2012ym} for more precise statements). This was partly resolved in \cite{Sannomiya:2016wlz}, showing various tacit assumptions in the theorem are violated, but the origin of the extra degrees of freedom remained a puzzle.}  We argue that the extra degrees of freedom originate from the spontaneous breaking of the emergent chiral supersymmetry in the continuum limit. As long as a linear dispersion relation appears, we do not need fine-tuning for the emergence. Moreover, the chiral supersymmetry is emanant in the free limit. More generally with interaction, it is almost emanant but strictly speaking it is only emergent in the continuum limit.

We also discuss the case when the spontaneously broken supercharge is invariant under the one-site lattice translation. In this case, there is no reason to expect the emergence of the chiral supersymmetry. At the same time, we cannot expect that the theory becomes relativistic in the low energy limit. Indeed, the dispersion relation of the Nambu-Goldstone mode becomes cubic. The chiral fermion number parity is emanant in the strict sense, and it does not act on the Nambu-Goldstone mode but it acts on the ``doubler" with a linear dispersion relation. 

With this non-relativistic model, one may be able to answer the question of whether the appearance of the doubled Nambu-Goldstone mode is a consequence of the doubler problem on the lattice. It is clearly related but it is not the necessary consequence. If we take the continuum limit first and formulate everything in terms of the continuum field without doubler problems, we still have doubled degrees of freedom in the relativistic case with the supersymmetry enhanced. In contrast, in the non-relativistic case, we can take the continuum limit by discarding the doubler with only one expected Nambu-Goldstone mode without enhanced supersymmetry.

The rest of the paper is organized as follows. In section 2, we study the emanant chiral supersymmetry in extended $\mathcal{N}=1$ Majorana Nicolai model. In section 3, we study the continuum version of the same theory. In section 4, we discuss its slight variant with cubic dispersion relation for a Nambu-Goldstone mode. In section 5, we study the emanant $\mathcal{N}=2$ chiral supersymmetry in the extended Nicolai model. The situation is more complicated and we see a couple of attempts to realize the emergent supersymmetries.

\section{Emanant chiral supersymmetry in extended $\mathcal{N}=1$ Majorana Nicolai model}

In this section, we study a Majorana fermion chain of length $L=2N$. we assume $N$ is a positive integer to define fermion number $(-1)^F$. 
At each site, we have a Majorana fermion operator $\gamma_j$ whose anti-commutation relation is given by the Clifford algebra $\{\gamma_i,\gamma_k\} = 2 \delta_{ij}$ with the reality condition $\gamma_i^\dagger = \gamma_i$.\footnote{With homage to Dirac we use the Greek letter $\gamma$ to represent Majorana fermion operators in this paper. (A string theorist might be against this use because $\gamma $ should be reserved for bosonic superghost.)} 

We introduce the vector supercharge
\begin{align}
Q^V = \sum_l^N \left( \frac{\gamma_{2l+1}}{2g} + \frac{i g}{2} \gamma_{2l-1} \gamma_{2l} \gamma_{2l+1} \right) \ . 
\end{align}
The notion of ``vector" here will be clarified once we introduce its axial version $Q^A$ below.
The corresponding Hamiltonian is 
\begin{align}
H &= \frac{1}{2} \{ Q^V, Q^V \}  = (Q^V)^2 \cr
&=  \sum_{l}^N \left( \frac{1}{4g^2} + \frac{i}{2} (\gamma_{2l} \gamma_{2l+1} + \gamma_{2l-1}\gamma_{2l} ) + \frac{g^2}{4}(1- 2\gamma_{2l-2}\gamma_{2l-1} \gamma_{2l+1} \gamma_{2l+2})  \right) \ .
\end{align}
By construction, the supercharge commutes with the Hamiltonian $[Q^V,H]=0$, and it is fermionic $(-1)^F Q^V (-1)^F = -Q^V$. This model is called the extended Majorana Nicolai model because Nicolai studied a similar model with complex fermion and without adding the linear term in the $\mathcal{N}=2$ supercharge \cite{Nicolai:1976xp}. The adjectives ``extended" and ``Majorana" refer to the two differences.

The phase structure of the ground state of $H$ has been worked out in the literature \cite{Sannomiya:2017foz}\cite{Miura:2023qph}\cite{Miura:2024hnh}\cite{Sannomiya:2024ffc}. There is a quantum critical point located at $g= g_c$. When $g<g_c$, the supersymmetry is spontaneously broken. The Nambu-Goldstone mode is dominated by $\gamma_{2l}$ (for small $g$), but the Nambu-Goldstone mode couples with the odd site fermion $\gamma_{2l+1}$ in the Hamiltonian, and they form the Majorana hopping term with a linear dispersion relation.\footnote{We would like to recall again that this linear dispersion is not predicted from the naive application of the results for the Nambu-Goldstone boson discussed in \cite{Watanabe:2012hr}\cite{Hidaka:2012ym} by replacing commutator with anti-commutator. Since $\langle 0| \{ Q^V, Q^V \} |0 \rangle \neq 0$, we could have expected the quadratic dispersion (type B) rather than the linear dispersion. Intuitively, in the bosonic theory, if the time derivative is first order, the spatial derivative must be second order to have a sensible theory, but in the fermionic theory, the first order spatial derivative may make sense as first observed by Dirac!} The continuum limit, therefore, is given by a free massless Majorana fermion (with irrelevant $T\bar{T}$ deformation), and the infrared central charge is $c=\frac{1}{2}$. This massless Majorana fermion is the continuum limit of the Nambu-Goldstone mode for the spontaneously broken supersymmetry $Q^V$. At $g=g_c$, supersymmetry is recovered in the continuum limit and the infrared theory is equivalent to the $\mathcal{N}=(1,1)$ supersymmetric minimal model with $c=\frac{7}{10}$ (or the fermionic version of the $(5,4)$ Virasoro minimal model \cite{Friedan:1984rv}\cite{Hsieh:2020uwb}, also known as the tri-critical Ising model).

It is a well-known story that the $\mathcal{N}=(1,1)$ supersymmetric minimal model has a supersymmetry preserving deformation that triggers the spontaneous supersymmetry breaking \cite{Friedan:1984rv}. The infrared theory is described by the massless Majorana fermion, but in the continuum limit, the broken supersymmetry is $\mathcal{N}=(1,1)$, and the massless Majorana fermion is the Nambu-Goldstone fermion for the $\mathcal{N}=(1,1)$ supersymmetry. We have two real supercharges, vector supersymmetry and axial supersymmetry that are both spontaneously broken. In contrast, in the lattice model that we started with, we have only one (exact)  vector supersymmetry that is spontaneously broken. Where is the other axial supersymmetry and what is the lattice origin of the twice Nambu-Goldstone mode that we observe in the continuum limit?

In the continuum limit, the axial supersymmetry and vector supersymmetry are related by the chiral fermion number parity transformation. To understand the lattice realization of the chiral fermion number parity, let us introduce the one-site lattice translation operator\footnote{An explicit expression of $T $ can be found e.g. in \cite{Sannomiya:2017foz}.} $T$ such that $T\gamma_{j}T^{-1} = \gamma_{j+1}$. The quadratic part of $H$ is invariant under $T$. As discussed in \cite{Seiberg:2023cdc}, it gives the emanant chiral fermion number parity $(-1)^{F_R}$ in the continuum limit. One salient feature of the lattice translation is that it anti-commutes (rather than commutes) with the fermion number operator $T(-1)^F T^{-1} = - (-1)^F$, suggesting the 't Hooft anomaly.\footnote{This can be easily seen by noting $(-1)^F$ is proportional to $\gamma_1 \gamma_2 \gamma_3 \cdots \gamma_{L}$, and crucially $L$ is even.}

The quadratic term of the Hamiltonian $H$ is invariant under the lattice translation by one unit, leading to the chiral fermion number parity symmetry of the free massless Majorana fermion.
The interaction term in the Hamiltonian, however, is not invariant under $T$. 

It is instructive to diagonalize the quadratic term of the Hamiltonian explicitly to see the action of $T$ and its relation to the chiral fermion number parity $(-1)^{F_R}$. Let us introduce the momentum space Majorana fermion operator
\begin{align}
\gamma_p = \sqrt{\frac{1}{N}} \sum_j^{2N} \gamma_j e^{-ipj}  \ ,
\end{align}
where $p = \frac{2\pi k}{N}$ (with integer $k$). Then the Hamiltonian can be expressed as
\begin{align}
H &= \text{const} + \sum_p \sin |p| \gamma_p^\dagger \gamma_p  + O(g^2) \cr
&\sim \text{const}   + \sum_p \left( |p| \gamma_p^\dagger \gamma_p + |p|\gamma_{\pi-p}^\dagger \gamma_{\pi-p} \right) \ .
\end{align}
Note that the dispersion relation has a doubler zero at $p=\pi$ in addition to $p=0$. The relevant action of $T$ is $T\gamma_{p=0}T^{-1} = \gamma_{p=0}$ and $T \gamma_{p=\pi} T^{-1} = - \gamma_{p = \pi}$ . The chiral fermion number parity acts non-trivially on the ``doubler" mode.  We also note that the Nambu-Goldstone mode for the spontaneously broken $Q^V$ is the sum of $\gamma_{p=0}$ and $\gamma_{p=\pi}$ because it is located only at the odd sites. Note also that in order to realize our supersymmetry, we have to keep the ``doubler" mode: we cannot discard the ``doubler" mode even in the continuum limit. See section 3.

To understand the origin of the extra massless degrees of freedom, let us define the axial supercharge $Q^A$ by
\begin{align}
Q^A = TQ^V T^{-1} = \sum_l^N  \left(\frac{\gamma_{2l}}{2g} + \frac{i g}{2} \gamma_{2l} \gamma_{2l+1} \gamma_{2l+2} \right) \ .
\end{align}
The anti-commutator gives 
\begin{align}
\tilde{H} &= T H T^{-1} = \frac{1}{2} \{ Q^A, Q^A \} = (Q^A)^2 \cr
&=  \sum_{l}^N \left( \frac{1}{4g^2} + \frac{i}{2} (\gamma_{2l+1} \gamma_{2l+2} + \gamma_{2l}\gamma_{2l+1} ) + \frac{g^2}{4}(1- 2\gamma_{2l-1}\gamma_{2l} \gamma_{2l+2} \gamma_{2l+3})  \right) \ .
\end{align}
Note that $\{ H , Q^A\} \neq 0 $ thus $[H, \tilde{H}] \neq 0$, but the spectrum of $H$ and $\tilde{H}$ is identical because they are related by the similarity transformation. There are two independent limits where the difference disappears. One limit is $g=0$, and the other is the continuum limit (even with non-zero $g$).

Thus we conclude that the axial supersymmetry and the chiral symmetry are emergent in the continuum limit with non-zero $g$. It is tempting to call it emanant chiral (or axial) supersymmetry but strictly speaking, it is conserved only in the continuum limit. The axial supersymmetry in the continuum limit is spontaneously broken and it explains the extra Majorana mode that is necessary to describe a relativistic massless Majorana fermion. The $g=0$ limit is special because one can regard the axial supersymmetry as an emanant symmetry (without taking the continuum limit). 

For completeness, let us compute
\begin{align}
\{Q^A, Q^V \} = -\frac{i}{2}\sum_l^N \left( \gamma_{2l} \gamma_{2l+2} + \gamma_{2l-1} \gamma_{2l+1} \right) = 2\tilde{P} \ . 
\end{align}
It ``generates" the lattice translation in the sense $[2i\tilde{P}, \gamma_i] = \gamma_{i+2} - \gamma_{i-2}$, but we would like to emphasize that this is not the symmetry of $H$ nor $\tilde{H}$ when $g\neq 0$. The two-site translation, which is the symmetry of the Hamiltonian, should have been realized as the unitary transformation $T^2 \chi_{i} T^{-2} = \chi_{i+2}$ instead.

There is a simple reason why we cannot preserve both $Q^A$ and $Q^V$ simultaneously on a lattice without taking the continuum limit as long as $g\neq 0$. If we were able to do so, the anti-commutator of the two should give a momentum generator, but the momentum generator cannot appear as a symmetry on a discrete spatial lattice. We cannot hope that the anti-commutator should give the finite lattice translation because the supersymmetry transformation is intrinsically infinitesimal.\footnote{Obviously, this has been a major difficulty in formulating supersymmetry on spatial (Euclidean) lattices, and this is one reason why we started with continuum time. One way to resolve the issue is to use the twisted supersymmetry whose anti-commutator gives on-site symmetry (or zero). See e.g. \cite{Catterall:2009it} for a review.}

In the $g=0$ limit, $\tilde{P}$ becomes a symmetry of the Hamiltonian, so we can regard $Q^A$ and the lattice chiral fermion number parity $(-1)^{F_{L}}$ are both emanant. Note that the $\tilde{P}$ is analogous to the Onsager algebra observed in emanant chiral $U(1)$ symmetries studied in \cite{Chatterjee:2024gje}. There, they realize that the emanant chiral $U(1)$ symmetry is preserved only in quadratic Hamiltonian and here we see a similar pattern: chiral supersymmetry is emanant only in the $g=0$ limit, otherwise it is emergent.

So far we have argued that the supercharges $Q^V$ and $Q^A$ cannot be a symmetry of the lattice Hamiltonian simultaneously. 
We, however, note that something special happens at $g=1$. While they are not the symmetry of the Hamiltonian, both of them annihilate the same ground states \cite{Sannomiya:2024ffc}! The trick is to rewrite the supercharges as
\begin{align}
Q^V &= \frac{1}{2}\sum_l^N \gamma_{2l-1} (1 + i \gamma_{2l} \gamma_{2l+1})  =  \frac{1}{2}\sum_l^N \gamma_{2l+1} (1 + i \gamma_{2l-1}\gamma_{2l} )  \cr
Q^A &= \frac{1}{2} \sum_l^N \gamma_{2l+2} (1 + i \gamma_{2l} \gamma_{2l+1}) =   \frac{1}{2}\sum_l^N \gamma_{2l-2} (1 + i \gamma_{2l-1}\gamma_{2l} )   \ . 
\end{align}
One can easily construct states $|\Phi_0 \rangle$ and $|\Phi_1\rangle$ such that $(1+i\gamma_{2l}\gamma_{2l+1} )|\Phi_0\rangle = 0 = (1+i\gamma_{2l-1} \gamma_{2l})|\Phi_1 \rangle$. Thus $|\Phi_0 \rangle$ and $|\Phi_1 \rangle$ are both annihilated by $Q^V$ and $Q^A$ simultaneously. Note that $T|\Phi_0 \rangle= |\Phi_1\rangle $ and $T$ can act non-trivially on the ground states. This is necessary because $T$ is anomalous and must respect the 't Hooft anomaly matching.

Since there is no superiority of $Q^V$ over $Q^A$, it may be an idea to define the total Hamiltonian as the average of $H$ and $\tilde{H}$. This approach was taken in \cite{OBrien:2017wmx}:
\begin{align}
H^{OF} &= \frac{1}{2}(H + \tilde{H}) \cr
& = \sum_j^{2N} \left( i\chi_{j} \chi_{j+1} - \frac{g^2}{2} \gamma_{j-2} \gamma_{j-1} \gamma_{j+1} \gamma_{j+2} \right) + \text{const} \ .
\end{align}
This Hamiltonian does not preserve the supersymmetry $Q^V$ or $Q^A$, but it is invariant under the one-site lattice translation $T$. In other words, it has a strict emanant chiral fermion number parity symmetry (even with nonzero $g$). As we have already discussed, we cannot realize any two out of the vector supersymmetry, axial supersymmetry, and the chiral fermion number parity simultaneously in an exact manner without taking the continuum limit (when $g\neq 0$). The spectrum and the phase structure of $H^{OF}$ were extensively studied in \cite{OBrien:2017wmx}. They are more or less the same as those of $H$ or $\tilde{H}$. 

\section{A continuum limit}
To demonstrate the claim that the axial supersymmetry and chiral fermion number parity emerge in the continuum limit of the lattice theory, we directly study the supercharge in the continuum limit.

Consider the supercharge
\begin{align}
Q^V = \int dx \left( \frac{\gamma}{2g} + \frac{ig}{2} \gamma \tilde{\gamma} \partial_x \gamma \right) \ . 
\end{align}
Here, $\gamma$ and $\tilde{\gamma}$ are independent Majorana fermion fields. We assume the canonical kinetic term so that $\{\gamma(x,t), \gamma(x',t) \} = 2\delta(x-x') = \{\tilde{\gamma}(x,t) , \tilde{\gamma}(x',t) \}$. The corresponding kinetic term in the action is $S_K = \int dx dt \frac{i}{4} (\gamma \partial_t {\gamma} + \tilde{\gamma} \partial_t  \tilde{\gamma})$.\footnote{If you are not familiar with the quantization of Majorana fermion, be aware of factor $2$.}

The Hamiltonian is given by
\begin{align}
H &= \frac{1}{2}\{Q^V,Q^V\} \cr
&= \int dx  \left( \frac{1}{4g^2} + \frac{i}{4}(\gamma \partial_x \tilde{\gamma} + \tilde{\gamma} \partial_x \gamma)  - \frac{g^2}{2} \partial_x (\gamma \tilde{\gamma}) \partial_x (\gamma \tilde{\gamma}) \right) \ . 
\end{align}
Note that the Hamiltonian is invariant under the exchange of $\gamma$ and $\tilde{\gamma}$. This will be the origin of the chiral fermion number parity.

We may rewrite the corresponding action in a more familiar form by introducing $\psi_L = \frac{\gamma+ \tilde{\gamma}}{2}$ and $\psi_R = \frac{\gamma-\tilde{\gamma}}{2}$.
\begin{align}
S = \int dx dt &\left(\frac{i}{2}(\psi_L \partial_t \psi_L + \psi_R \partial_t \psi_R) -\frac{i}{2}(\psi_L \partial_x \psi_L - \psi_R \partial_x \psi_R)  \right. \cr 
  & \left. -\frac{1}{4g^2}+ 2g^2\partial_x(\psi_L \psi_R) \partial_x(\psi_L \psi_R)   \right) \ . 
\end{align}
The kinetic term of the action is manifestly Lorentz invariant and it describes a (massless) Majorana fermion of the left chirality and the right chirality. The interaction term is not manifestly Lorentz invariant, but with the non-linear field redefinition, it is possible to recast in the form of the Lorentz invariant $T\bar{T}$ deformation \cite{Lee:2021iut,Smirnov:2016lqw,Cavaglia:2016oda} (at least perturbatively in $g$). The price we had to pay to obtain the Lorentz invariant interaction term is that the anti-commutation relation becomes non-standard (i.e. field dependent). In the lattice formulation, it is more convenient to keep the standard anti-commutation relations. The exchange symmetry of $\gamma$ and $\tilde{\gamma}$ flips only the sign of $\psi_R$, so it is indeed the chiral fermion number parity.

We, therefore, claim that as long as $g \le g_c$, the continuum theory is Lorentz invariant. At $g=g_c$, we will have a fermionic description of the $(5,4)$ minimal model. It is not, however, obvious if the theory remains Lorentz invariant when $g>g_c$. We here give an argument that it cannot be Lorentz invariant at the exact supersymmetric point (i.e. $g=1$ in the lattice theory). The point is that the vacuum there is characterized by demanding the supercharge density annihilates the vacuum $q^V(x)|0\rangle = \left(\frac{\gamma}{2} + \frac{i}{2} \gamma \tilde{\gamma} \partial_x \gamma\right) |0\rangle = 0$ without integration over $x$. Suppose the continuum theory is relativistic. Then the Reeh–Schlieder theorem claims that $q^V(x)$ must be zero i.e. the vacuum is a separating vector. It would mean that the Hamiltonian must be identically zero, so the non-trivial continuum theory cannot be relativistic.\footnote{The same argument applies to any frustration-free systems, where a local operator annihilates the vacuum. The Reeh-Schlieder theorem tells us that frustration-free systems cannot be relativistic. See \cite{Masaoka:2024oes} for a different perspective on this.}

We now introduce the axial supercharge 
\begin{align}
Q^A = \int dx \left( \frac{\tilde{\gamma}}{2g} + \frac{ig}{2} \tilde{\gamma} {\gamma} \partial_x \tilde{\gamma} \right) \ . 
\end{align}
It gives the same Hamiltonian 
\begin{align}
\tilde{H} =  \frac{1}{2} \{Q^A,Q^A\} = H \  
\end{align}
so it commutes with the (original) Hamiltonian $[H, Q^A] = 0$. As emphasized earlier, the continuum theory has two supersymmetry $Q^V$ and $Q^A$ at the same time. To check the closure of the algebra, let us compute 
\begin{align}
\{Q^A, Q^V \} = -\int dx\frac{i}{2}  \left( \tilde{\gamma} \partial_x \tilde{\gamma} + \gamma \partial_x {\gamma}  \right) = 2P \  . 
\end{align}
Here, $[P,H] = 0$ and it generates the spatial translation. In total we have $\mathcal{N}=(1,1)$ supersymmetry algebra. 
When $g<g_c$, supersymmetries are spontaneously broken and they give one massless Majorana fermion as a Nambu-Goldstone mode, but this is all expected and we have no counting issues here without taking the $g=0$ limit. At the same time, we want to emphasize that in the continuum limit, we have kept the doubler mode to realize non-trivial supersymmetry with interaction.

\section{Non-relativistic $\mathcal{N}=1$ supersymmetry}
The supercharges used in previous sections are not invariant under the one-site lattice translation. It is possible to find a supercharge invariant under the one-site lattice translation. Let us consider the real supercharge \cite{Sannomiya:2017foz}:
\begin{align}
Q^C = \sum_j^{2N} \left(\frac{\gamma_j}{g} + i g \gamma_j \gamma_{j+1} \gamma_{j+2} \right) \ , 
\end{align}
which satisfies $T Q^c T^{-1} = Q^c$. Accordingly, the Hamiltonian 
\begin{align}
H^C &= \frac{1}{2} \{ Q^C, Q^C \} = (Q^C)^2 \cr
& = \sum_j^{2N} \left(\frac{1}{g^2} + 2i (2\gamma_j \gamma_{j+1} - \gamma_{j-1} \gamma_{j+1}) + g^2 \sum_j (1- 2\gamma_{j-2}\gamma_{j-1} \gamma_{j+1}\gamma_{j+2}) \right)
\end{align}
is invariant under the one-site lattice translation.

This Hamiltonian shows the spontaneous supersymmetry breaking when $g<g_c$. Once the supersymmetry is spontaneously broken, the low energy mode contains the Nambu-Goldstone mode with a cubic dispersion relation and one extra mode with a linear dispersion relation. The latter is not associated with the spontaneous supersymmetry breaking but it is associated with the ``doubler problem" in lattice fermion: the dispersion must be periodic in $p$ but if we have only one cubic mode, how can we make the dispersion periodic? In this sense, the counting rule of the Nambu-Goldstone mode is not violated and there is no emergent supersymmetry in this model.\footnote{The dispersion relation, however, is exceptional. See e.g. for more details on this point.}
This theory has an emanant chiral fermion number parity, which acts on the linear mode. This ``doubler" should have a linear dispersion relation rather than the cubic dispersion relation so that it is charged under the emanant chiral fermion number parity.

To see these statements more explicitly, let us diagonalize the quadratic terms of the Hamiltonian by introducing the momentum space Majorana fermion operator
\begin{align}
\gamma_p = \sqrt{\frac{1}{2N}}\sum_{j}^N \gamma_j e^{-ipj}
\end{align}
so that 
\begin{align}
 H &= \text{const} + \sum_{ 0\le p \le \pi} (2 \sin(p) - \sin(2p) \gamma^\dagger_p \gamma_p  + O(g^2) \cr
 &\sim \text{const} + \sum_{p > 0} \left(|p|^3 \gamma^\dagger_p \gamma_p + |p| \gamma^\dagger_{\pi -p} \gamma_{\pi -p} \right) \ .
\end{align}
There is an emanant chiral fermion number parity symmetry that acts on the linear mode. Note that the Nambu-Goldstone mode is $\gamma_{p=0}$ and does not mix with $\gamma_{p=\pi}$ in stark contrast with the relativistic case studied in section 2. It has a cubic dispersion relation and is not charged under the chiral fermion number parity. The chiral fermion number parity acts non-trivially only on the mode around $p=\pi$. 

Let us briefly study its continuum limit. Here, we can formulate the continuum limit without introducing the doubler mode. Let us introduce one Majorana field $\gamma(x)$. The supercharge is defined as
\begin{align}
Q^C = \int dx \left( \frac{\gamma}{g} +i g \gamma \partial_x \gamma \partial_x^2 \gamma \right) \ . 
\end{align}
The continuum Hamiltonian is given by
\begin{align}
H^C = \frac{1}{2} \{ Q^C, Q^C \} = \int dx \left( \frac{1}{g^2} + 2i\partial_x\gamma \partial^2_x \gamma + 8g^2 \gamma \partial_x \gamma \partial_x^2 \gamma \partial_x^3 \gamma \right) \ . 
\end{align}
When $g<g_c$, the supersymmetry $Q^C$ is spontaneously broken and the Nambu-Goldstone mode has a cubic dispersion relation $E\propto |p|^3$. Note that there is no chiral fermion number parity symmetry in this continuum limit. Indeed, the ``doubler" mode with linear dispersion relation is eliminated.\footnote{In other words, if we want to realize this particular non-relativistic theory with cubic dispersion on a lattice, the lattice regularization in this section may not be satisfactory because we have to introduce the ``doubler" mode. } 

\section{Emanant $\mathcal{N}=2$ Supersymmetry?}

In this section, we attempt to uncover the doubled Nambu-Goldstone modes in the lattice fermion model with the spontaneously broken $\mathcal{N}=2$ supersymmetry studied in \cite{Sannomiya:2017foz}. We start with the fermionic chain with Dirac (or complex) fermion operators $c_i$ at each site. The anti-commutation relation is $\{c_i, c_j\} = 0 = \{c_i^\dagger, c_j^\dagger \}$ and $\{c_i, c_j^\dagger\} = \delta_{ij}$. 

The complex supercharge studied in \cite{Sannomiya:2017foz} is
\begin{align}
Q_V = \sum_k^N \left( \frac{c_{2k-1}}{g} + g c_{2k-1} c_{2k}^\dagger c_{2k+1} \right) \ . 
\end{align}
Note that $\{Q_V,Q_V\} = 0$. The Hamiltonian is given by 
\begin{align}
H &= \{Q_V, Q_V^\dagger\} \cr
&= \frac{N}{g^2} + \sum_j^{2N} (-1)^j (c_j^\dagger c_{j+1} + c_{j+1}^\dagger c_j) + g^2 H_{\text{int}}
\end{align}
where the four-fermi interaction $H_{\text{int}}$ can be found in \cite{Sannomiya:2017foz}. Since $\{ Q_V, Q_V\} = 0$, the supercharge $Q_V$ commutes with the Hamiltonian.
The supersymmetry $Q_V$ is spontaneously broken for finite $g$ on any finite lattice.

In the continuum limit, it is described by the massless Dirac fermion (with Thirring interaction of order $g^2$ \cite{Sannomiya:2017foz}). This massless Dirac fermion is the Nambu-Goldstone mode, but as in the $\mathcal{N}=1$ case above, the relativistic massless Dirac fermion has twice the degrees of freedom than that for the broken supersymmetry. Indeed, $c^\dagger_{2k-1}$ creates the Nambu-Goldstone mode, but $c^\dagger_{2k}$ also creates the same mode due to the kinetic mixing!

To look for the origin of the twice degrees of freedom in the Nambu-Goldstone mode, we now discuss possible emanant or emergent supersymmetry. For this purpose, it may be worthwhile recalling the lattice chiral fermion number parity and the lattice $U(1)_A$ symmetry. The lattice chiral fermion number parity is related to the one-site lattice translation by $c_{i} \to c_{i+1}$. As discussed in \cite{Chatterjee:2024gje}, the lattice $U(1)_A$ symmetry may be realized as $ \frac{1}{2} \sum_j(c_j^\dagger + c_{j})(c_{j+1}^\dagger - c_{j+1})$ or other possibilities e.g. in \cite{Banks:1975gq}. This lattice $U(1)_A$ is not invariant under $U(1)_V$ symmetry, and we will see a similar structure in the following.

Motivated by the one-site lattice translation, let us introduce the axial supercharge 
\begin{align}
Q_A = \sum_k^N  \left(\frac{c_{2k}}{g} - g c_{2k} c_{2k+1}^\dagger c_{2k+2} \right) \ . 
\end{align}
Here, $\{Q_A,Q_A\} = 0$ and 
\begin{align}
\tilde{H} &= \{Q_A, Q_A^\dagger\} \cr
&= \frac{N}{g^2} + \sum_j (-1)^j (c_j^\dagger c_{j+1} + c_{j+1}^\dagger c_j) + g^2 \tilde{H}_{\text{int}} \ . 
\end{align}
$\tilde{H}_{\text{int}} \neq H_{\text{int}} $ but more importantly the difference does not vanish even in the continuum limit. This is the first difference of the $\mathcal{N}=2$ supersymmetric case than in the $\mathcal{N}=1$ supersymmetric case discussed in sections 2 and 3. There, the situation was rather fortuitous because the leading irrelevant deformation was the $T\bar{T}$ deformation and was unique. The difference between $H$ and $\tilde{H}$ did vanish in the continuum limit even with the interaction terms.
Here we have a lot more choices. 

An additional difficulty is that the anti-commutator $\{Q_A, Q_V\}$ is non-zero but gives 
\begin{align}
\{Q_A, Q_V\} = \sum_{k}^N \left( c_{2k}c_{2k+2} - c_{2k-1} c_{2k+1} \right) + O(g^2) \ .
\end{align}
On the other hand, the anti-commutator $\{Q_A^\dagger, Q_V\}$ is zero while we could have expected that it is related to the two-sites lattice translation (which could emanate the momentum operator in the continuum limit).

We can still argue that $Q_A$ is an emanant chiral supersymmetry in the $g=0$ limit, but it is not satisfactory that the anti-commutator did not give the desired supersymmetry algebra. To realize the $\mathcal{N}=(2,2)$ supersymmetry and their spontaneous breaking more naturally on the lattice, it may be possible to use a similar idea to the ``averaged" Hamiltonian. Let us consider a complex ``supercharge":
\begin{align}
\mathcal{Q}_V = \sum_k^N \left( \frac{c_{2k-1}}{g} -  g (c^\dagger_{2k+3}-c^\dagger_{2k+1}) c_{2k-1} c_{2k+2} \right)  \cr
\mathcal{Q}_V^\dagger = \sum_k^N \left( \frac{c^\dagger_{2k-1}}{g} -  g  c_{2k+2}^\dagger c_{2k-1}^\dagger (c_{2k+3}-c_{2k+1})  \right) \label{ccv}
\end{align}
Unlike $Q_V$, the new supercharge $\mathcal{Q}_V$ does not generate the exact supersymmetry because $\{ \mathcal{Q}_V, \mathcal{Q}_V \} \neq 0$ but it is  order $O(g^2)$. Nevertheless, let us define the Hamiltonian by 
\begin{align}
\mathcal{H} &= \{ \mathcal{Q}_V, \mathcal{Q}_V^\dagger \} \cr
&= \sum_k^N  \left( \frac{1}{g^2} + c_{2k}^\dagger(c_{2k+1}-c_{2k-1}) + (c^\dagger_{2k+1} - c_{2k-1}^\dagger)c_{2k} \right) + O(g^2) \ .
\end{align}
Note that $H = \mathcal{H} + O(g^2)$. Since  $\{ \mathcal{Q}_V, \mathcal{Q}_V \} \neq 0$, the new supercharge $\mathcal{Q}_V$ does not commute with $\mathcal{H}$ in the strict sense $[\mathcal{H},\mathcal{Q}_V] = O(g^2)$. Nevertheless, some of the main properties of the supersymmetry such as that the zero energy ground states of $\mathcal{H}$ are annihilated by $\mathcal{Q}_V$ and $\mathcal{Q}_V^\dagger$ remain.

Nevertheless, merit of considering $\mathcal{Q}_V$ and $\mathcal{H}$ exists. We now introduce the axial version:
\begin{align}
\mathcal{Q}_A = \sum_k^N \left( \frac{c_{2k}}{g} +  g (c^\dagger_{2k+2}-c^\dagger_{2k}) c_{2k-2} c_{2k+1} \right) \cr
\mathcal{Q}_A^\dagger = \sum_k^N \left( \frac{c^\dagger_{2k}}{g} +  g  c_{2k+1}^\dagger c_{2k-2}^\dagger (c_{2k+2}-c_{2k}) \right) \label{cca}
\end{align}
If we define $\tilde{\mathcal{H}} = \{ \mathcal{Q}_A, \mathcal{Q}_A^\dagger \}$, we have $H = \tilde{\mathcal{H}} + O(g^2)$ again. However, one good news now is we have $\{\mathcal{Q}^A, \mathcal{Q}^V\} = 0$ exactly. Furthermore, 
\begin{align}
\{ \mathcal{Q}_A, \mathcal{Q}_V^\dagger \} = \sum_k^N  \left((c^\dagger_{2k+2} - c_{2k}^\dagger)c_{2k-2} - c^\dagger_{2k-1} (c_{2k+3} - c_{2k+1})  \right) + O(g^2) = \tilde{\mathcal{P}}  \ . 
\end{align}
It ``generates" a lattice analogue of the spatial translation.
So, if we neglect the interaction terms of order $g^2$, the expected (anti)-commutation relations can be realized on the lattice that becomes $\mathcal{N}=(2,2)$ supersymmetry in the continuum limit. The chiral $\mathcal{N}=(2,2)$ supersymmetries can be emanant in the non-interacting limit of $g=0$.

In this theory, it is believed that there is no phase transition as a function of $g$ at least in the continuum limit. However, the ground states will be exponentially degenerate in the strict $g=\infty$ limit.
It is fun to look at the number of ground state degeneracies when  $g=\infty$. The exact diagonalization shows that it is $(4),(6),26,58,138,344,876,2258, 5858 \cdots$ for $L=2,4,6, 8 \cdots$. It seems to be related to $4+$ ``A124697: Number of base 4 circular n-digit numbers with adjacent digits differing by 1 or less." which can be found in the integer sequence Wikipedia (aside from the first two). On that page, they show that the generating function is $A(x) = \frac{(1-3x^2-4x^4+3x^4)}{(1-3x + x^2)(1 - x - x^2)}$. 
The explanation of the number seems to suggest there may exist a classical counting reasoning behind such as the one that was discussed in \cite{Katsura:2017pjd}, or actually a much simpler one. It would be curious to see whether we can derive this from the technique used in \cite{La:2018zkj}.

Let us briefly mention the continuum limit at the level of the supercharges. We introduce the two complex fermion fields $c$ and $\tilde{c}$ with the anti-commutation relations: $\{c(x,t),c^\dagger(y,t)\} = \delta(x-y) = \{\tilde{c}(x,t), \tilde{c}^\dagger(y,t)\}$ (every others are anti-commuting at equal time). 

We begin with the continuum supercharges of the original extended Nicolai model \cite{Sannomiya:2016wlz}
\begin{align}
Q_V &= \int dx \left( \frac{c}{g} + c \tilde{c}^\dagger \partial_x c \right) \cr
Q_V^\dagger&= \int dx \left( \frac{c^\dagger}{g} + \partial_x c^\dagger \tilde{c} c^\dagger \right) \ . 
\end{align}
Since $\{Q_V, Q_V\}=0$, we can construct the supersymmetric Hamiltonian as
\begin{align}
H = \{Q_V,Q_V^\dagger\} = \int dx \left( \frac{1}{g^2} + \partial_x c^\dagger \tilde{c} + \tilde{c}^\dagger \partial_x c \right) + O(g^2) \ . 
\end{align}
The supersymmetry is spontaneously broken and the low energy dynamics with small $g$ is given by the free massless Dirac fermion with $c=1$.

We may define the axial supercharge as
\begin{align}
Q_A &= \int dx \left( \frac{\tilde{c}}{g} -\tilde{c} {c}^\dagger \partial_x \tilde{c} \right) \cr
Q_A^\dagger &= \int dx  \left( \frac{\tilde{c}^\dagger}{g} - \partial_x \tilde{c}^\dagger {c} \tilde{c}^\dagger \right) \ . 
\end{align}
so that 
\begin{align}
\tilde{H} = \{Q_A,Q_A^\dagger\} = \int dx \left( \frac{1}{g^2} - \partial_x \tilde{c}^\dagger  {c} - {c}^\dagger \partial_x \tilde{c} \right) + O(g^2)
\end{align}
We have $H = \tilde{H} + O(g^2)$. This is a candidate for the emergent supersymmetry that is spontaneously broken (in the $g=0$ limit).
As we have mentioned above, this axial supercharge does not generate the usual $\mathcal{N}=(2,2)$ supersymmetry algebra. To see this, let us compute the other anti-commutation relations:
\begin{align}
\{Q_V, Q_A \} &= \int dx  \left( \tilde{c} \partial_x \tilde{c} - c \partial_x c \right) + O(g^2) \cr
\{Q_V, Q_A^\dagger \} &= 0 \ . 
\end{align}

Finally, let us study the continuum theory of the ``averaged" version with non-conserved supercharges corresponding to \eqref{ccv} and \eqref{cca} (for finite $g$). Let us introduce
\begin{align}
\mathcal{Q}_V &= \int dx \left( \frac{c}{g} - g(\partial_x c^\dagger) c \tilde{c} \right) \cr
\mathcal{Q}_V^\dagger &= \int dx \left( \frac{c^\dagger}{g} - \tilde{c}^\dagger c^\dagger \partial_x c \right)  \cr
\mathcal{Q}_A &= \int dx \left( \frac{\tilde{c}}{g} + g (\partial_x \tilde{c}^\dagger) \tilde{c} c \right) \cr
\mathcal{Q}_A^\dagger & = \int dx \left( \frac{\tilde{c}^\dagger}{g} + c^\dagger \tilde{c}^\dagger \partial_x \tilde{c} \right) \ . 
\end{align}
and we find 
\begin{align}
\mathcal{H} &= \{\mathcal{Q}_V, \mathcal{Q}_V^\dagger \} = \int dx \left( \frac{1}{g^2} +\tilde{c}^\dagger \partial_x c + (\partial_x c^\dagger) \tilde{c} \right) + O(g^2) \cr
\tilde{\mathcal{H}} &= \{\mathcal{Q}_A, \mathcal{Q}_A^\dagger \} = \int dx \left( \frac{1}{g^2} -(\partial_x \tilde{c}^\dagger) c - c^\dagger \partial_x \tilde{c} \right) + O(g^2) 
\end{align}
Furthermore
\begin{align}
\{ \mathcal{Q}_V, \mathcal{Q}_A \} &= 0 \cr
\{\mathcal{Q}_V, \mathcal{Q}_A^\dagger \} &= \int dx \left( - \tilde{c}^\dagger \partial_x \tilde{c} + (\partial_x c^\dagger) c \right) + O(g^2) \ . 
\end{align}
They give the $\mathcal{N} = (2,2) $ supersymmetry in the $g=0$ limit. In particular, $\{ \mathcal{Q}_V, \mathcal{Q}_A^\dagger \} = 2P$ becomes the momentum generator.
Thus, we verify that the $\mathcal{N}=(2,2) $ supersymmetry is emanant in the $g=0$ limit, and we have a massless Dirac fermion as a Nambu-Goldstone mode. It should be extremely interesting to find a model with the emergent (or more hopefully emanant) $\mathcal{N}=(2,2)$ supersymmetry without taking the $g=0$ limit.

\section{Conclusion}
Counting fermionic Nambu-Goldstone mode is much more non-trivial than bosonic cases in particular on a finite lattice. One issue is how to realize the symmetry exactly on a lattice and the other issue is the ``doubler" problem. 
The extended Majorana Nicolai model has a distinguished feature that the supersymmetry is spontaneously broken but the counting rule of the Nambu-Goldstone mode is exceptional. We have found the origin is associated with hidden emanant or emergent supersymmetry. At the more technical level, as discussed in \cite{Sannomiya:2016wlz}, the assumption of the analyticity in \cite{Nielsen:1975hm} can be violated e.g. $E = |p|$ vs $E^2 = p^2$ with the fermi sea, and the counting should be often modified further.

There are several other interesting examples where supersymmetry breaking can be realized in lattice systems. One interesting example is the two-dimensional lattice QCD with massive adjoint fermions \cite{Dempsey:2023fvm}\cite{Dempsey:2024alw}. At a particular mass, the theory becomes supersymmetric and there exists a vacua with a spontaneously broken supersymmetry. It will be interesting to study how the (spontaneously broken) supersymmetry is realized on a finite lattice. Another interesting direction is a higher dimensional generalization. See e.g. \cite{Affleck:2017ubr}\cite{Wamer:2018fpt}\cite{Li:2018lxd} for some attempts to realize (spontaneously broken) supersymmetry by using only fermions.

\section*{Acknowledgments}
The author is in part supported by JSPS KAKENHI Grant Number 21K03581. The results in sections 2 and 3 were presented at Yukawa Institute for Theoretical Physics at Kyoto University in January 2024 and Princeton Center for Theoretical Science at Princeton University in February 2025. He would like to thank the participants for the stimulating atmosphere and inspiring feedback. He, in particular, thanks Igor Klebanov for his hospitality during the visit.

\end{document}